\documentclass[runningheads,a4paper]{llncs}

\usepackage{amssymb}  \usepackage{graphicx}
\usepackage{cite}

\usepackage[tophrase=--]{siunitx}
\usepackage{subfig}
\usepackage{calc}
\begin{document}

\mainmatter% start of an individual contribution

\titlerunning{Combining Brain-Computer Interfaces and Haptics}
\title{Combining Brain-Computer Interfaces and Haptics: Detecting Mental Workload to Adapt Haptic Assistance}

\authorrunning{L. George, M. Marchal, L. Glondu and A. L\'ecuyer}
\author{Laurent George\inst{1,2,4}, Maud Marchal\inst{1,2,4}, Loeiz Glondu\inst{3,4},
and~Anatole~L\'ecuyer\inst{1,4}% <-this % stops a space
}
% the affiliations are given next; don't give your e-mail address unless you accept that
% it will be published
\institute{INRIA, Rennes, France
\and
INSA, Rennes, France
\and
ENS Cachan, Bruz, France
\and
IRISA, Rennes, France
}
\maketitle
\sloppy
%%% PATH FOLLOWING TASK partout et pas precision tracking ..
\begin{abstract}
  In this paper we introduce the combined use of Brain-Computer Interfaces (BCI) and Haptic interfaces. We propose to adapt haptic guides based on the mental activity measured by a BCI system. This novel approach is illustrated within a proof-of-concept system: haptic guides are toggled during a path-following task thanks to a mental workload index provided by a BCI. The aim of this system is to provide haptic assistance only when the user's brain activity reflects a high mental workload.
  A user study conducted with 8 participants shows that our proof-of-concept is operational and exploitable. Results show that activation of haptic guides occurs in the most difficult part of the path-following task. Moreover it allows to increase task performance by 53\% by activating assistance only 59\% of the time. Taken together, these results suggest that BCI could be used to determine when the user needs assistance during haptic interaction and to enable haptic guides accordingly.
  %and suggest that enabling assistance based on measured brain activity makes it possible to reduce mental workload and improve task performance.
  % donner des chiffres tend to confirm the operability of such a system.  to assess an online index related to user's mental workload. When the task leads to a high mental workload (e.g. during precision movements) the assistance was provided. Preliminary results with eight participants showed that this proof-of-concept system is operational and allow to reduce user mental workload and to improve performance by reducing number of collisions. 
\end{abstract}

\section{Introduction} A Brain-Computer Interface (BCI) is a communication system that transfers information directly from brain activity to a computerized system~\cite{wolpaw2002brain}. The original goal of BCI is to provide control and communication capabilities for people with severe disabilities. For example, BCI-based spellers enable to spell letters by only using brain activity~\cite{farwell}. Another example is the use of BCI to send commands to prostheses, for example to open and close prostheses by imagination of hand movements~\cite{Guger}. BCI can also be used by healthy users, for instance for enhancing interaction with video games~\cite{healthy}. An approach called ``passive BCI'' aims at using brain activity information to adapt and enhance the current application without the need for the user to voluntarily control his/her brain activity~\cite{George2010,Zander2011a}. For example, this approach has been used to adapt virtual environments content~\cite{Nijholt2009}. Adapting interaction modalities according to the user mental state could also be another interesting option offered by passive BCIs.

 Haptic feedback has already been used in a BCI system~\cite{Muller-Putz2006,Cincotti2007,Chatterjee2007}. However, to our best knowledge, BCI have never been used for adapting haptic feedback yet. In this paper, we introduce the use of the passive BCI approach in the haptic realm.

We propose to use BCI technologies to adapt force-feedback in real-time. We introduce assistive tools, i.e.\ haptic guides, which are automatically and continuously adapted to the user's mental workload measured through a passive BCI.

The remainder of this paper is organized as follows. Related work is presented in Section~\ref{sec:relatedwork}. In Section~\ref{sec:concept} we detail our concept of haptic assistance based on BCI. An implementation of this concept and a user study are proposed in Section~\ref{sec:method}. We discuss results in Section~\ref{sec:discussion}. Finally the main conclusions are summarized in Section~\ref{sec:conclusion}.

\section{Related Work}
\label{sec:relatedwork}
% - BCI passive
Our description of related work is subdivided into three parts. First we provide a brief summary of previous work on haptic guidance. Second we present related work in the field of BCI, notably the use of BCI in virtual environments and the passive BCI approach. In the last part we give an overview of how haptic feedback and BCI have already been combined together.

\subsection{Haptic guidance}
%% TODO : URGENT : faire definition haptic guidance
Haptic guidance can be defined as an interaction paradigm in which the user is physically guided through an ideal motion by a haptic interface~\cite{Feygin2002}. Bluteau~et~al.~\cite{Bluteau2008} have compared different types of guidance and showed that the addition of haptic information plays an important role on the visuo-manual tracking of new trajectories, especially when forces are used for the guidance. Recent work has focused on the adaptation of haptic guidance to maximize learning effect~\cite{Asseldonk2009}.  

\subsection{BCI}
\paragraph{BCI and Virtual Environments:}
BCI have been early demonstrated as usable for interacting within 3D virtual environments~\cite{Lecuyer2008}. They have been notably used for navigating in the virtual environment or moving a virtual object~\cite{Lecuyer2008}. For a comprehensive survey on the combintation of BCI and Virtual Reality and videogames the interested reader can refer to recent surveys~\cite{Lecuyer2008, Nijholt2009}. % TODO anatole pensais a un troisieme nihjohlt
%Virtual environment have also been used to test BCI in a safe and controllable environment. %For example, Leeb~et~al.~\cite{Leeb2007} proposed to use a virtual environment simulation for testing a wheelchair control based on BCI\@. 
 BCI have also been used in novel interaction paradigms, where the BCI would not substitute for classical interaction peripherals but complement them. For instance it has been proposed to use the brain activity information for changing the interaction protocol or the content of the virtual world~\cite{Lecuyer2008}, which refers to the ``passive BCI'' approach.

 \paragraph{Passive BCI:} The aim of a passive BCI is not to allow the user to send explicit commands to an application but to provide information concerning his/her mental state with the purpose of adapting or enhancing the interaction accordingly~\cite{George2010,Zander2011a}. For instance, passive BCI have been used in the context of videogames to adapt the way the system responds to commands~\cite{Muhl2010} or to adapt the content of the game itself~\cite{Nijholt2009}. In ``Alpha WoW''~\cite{Nijholt2009} the user avatar form is updated (from elf to wolf) according to the measured level of alpha activity. The aim is to provide the most adequate avatar according to the context (e.g.\ a detected stress would activate an avatar meant for close-combat). Passive BCI approach has also been used in real driving environment~\cite{Kohlmorgen2007}. In their system, Kohlmorgen et al.\ proposed to use passive BCI to interrupt secondary tasks of a driver when detecting a high mental workload~\cite{Kohlmorgen2007}.

%This approach seems particularly relevant in the context of virtual %environment where the user could benefits of a more user aware system.  In the %adaptive automation field Pope et al.\ proposed a system that adapt the level %of automation of a tracking task using an engagement index measure throug %Electroencephalography~\cite{Pope1995} (e.g a loss of engagement leads to the %diminution of automation). Passive BCI have also been used with the aim to %correct errors. Indeed error related potentential which are reactions of the %brain to error can be measured by BCI system~\cite{Ferrez2007}.  %In~\cite{Parra2003}, Parra et al. use the detection of error potentials to %correct errors in a visual discrimination task. In~\cite{Zander2011}, Zander %et al.\ proposed to use a passive BCI system to detect perception of correct %and erroneous ending of sequences of chords.

\subsection{Haptic for BCI}
% - BCI et tactile 
Haptic interfaces have already been combined with BCI for providing stimuli and feedback for BCI systems.  
Vibro-tactile stimuli have notably been demonstrated to be usable for BCI systems. M\"uller-Putz~et~al.~\cite{Muller-Putz2006} showed that using vibro-tactile stimuli on fingertips results in brain activity that can be modulated by the user will. Focusing on left-hand or right-hand finger stimulation allows the user to send commands using what is called a ``Steady-State Somatosensory Evoked Potential''~\cite{Muller-Putz2006}. % TODO voir avec Anatole si c'est la ref ? ou s'il veut une ref sur somatosensory
% - BCI et force-feedback
Force-feedback has also been used to provide valuable feedback to the user during BCI interaction~\cite{Cincotti2007,Chatterjee2007}.  

However, to the authors' best knowledge, there has been no previous work on the adaptation of haptic feedback thanks to a BCI, using a passive BCI approach.

\section{Using mental workload to adapt haptic assistance}
\label{sec:concept}
% bien découper : 
% premier parragraphe : dire ce qu'on veut faire, donner les hypotheses
The concept proposed in this paper consists in using a passive BCI to assess in real-time an index of the user's mental workload and to adapt haptic assistance accordingly. 

%%% TODO : revoir les exemples ici 
\emph{Mental workload} is a generic term which can cover or apply to different and varied cognitive processes and mental states. It could apply for example to a memorization task (e.g.\ image memorization), driving task, lecture task and cognitive task~\cite{heger2010}. In this paper we use the term mental workload to qualify the modification of the user brain activity in relation to the difficulty of a manipulation task. Mental workload index is expected to increase with the manipulation difficulty level. Several EEG markers have been identified as correlated with mental workload, task engagement or attention~\cite{Pope1995,Kohlmorgen2007,heger2010}. In~\cite{Pope1995} the authors proposed to use ratios of activity in specific band-power such as alpha (\SIrange{8}{12}{\hertz}) or theta (\SIrange{3}{7}{\hertz}) bands to compute an index of the user task engagement. More recently, more complex approaches based on machine learning have been used to compute index of user's mental workload based on EEG activity~\cite{Kohlmorgen2007,heger2010}. In this paper, we propose to use a BCI system based on these approaches for assessing an index of the user's mental workload during a haptic interaction task.

%We can also note that user mental workload knowledge could be used during the %design and evaluation of a haptic simulation (e.g\ knowing which part of a %simulation elicit a high mental workload for refining the simulation).

\emph{Haptic assistance systems} could benefit from the user mental workload information. Indeed, it could be used to know when the user needs assistance or not. For example, a high mental workload could present a risk in the context of safety-critical haptic tasks. A smart assistance system that interrupts the haptic interaction or toggles specific haptic guides might improve comfort or safety of operations in such conditions. Haptic guides that would be only active when the user presents a high mental workload (e.g.\ the user is focused on a difficult precision gesture) might improve task performances and learning process.
%
% Knowing in real time this information seems to be particularly useful in learning simulation. Indeed, this information could be used to ensure that the user stays at a specific mental workload level. Thus for providing a challenging experience, we could imagine a system that adapts the difficulty level of the simulation based on the user mental workload (e.g.\ increasing the difficulty until a specific level is reached). Mental workload information could also benefit to an intelligent assistive system. Indeed mental workload assessment could help to know which part of a simulation present difficulty for a user; assistive system can then be launched accordingly.
%% passive BCI -> nous permet d'obtenir le workload
%This kind of system may also be used during safety-critical tasks. Indeed a too %high mental workload could present risk in this context. An intelligent system %that interrupts the haptic interaction, raises an alert and activates haptic %guiding assistance may help to improve safeness of operation in such condition.
%% et maintenant nous on regarde le cas haptique

\section{Evaluation}
\label{sec:method}
Thereafter, we design and evaluate a proof-of-concept system that toggles a haptic guidance during a path-following task based on the user mental workload. The task consists in following a path in a virtual 2D maze avoiding collisions with borders. An EEG-based BCI is used to compute an online index related to the user mental workload. If the index indicates a high (resp.\ low) mental workload, the haptic assistance is activated (resp.\ deactivated). 

\subsection{Objectives and hypotheses}
% TODO : objs : opérationnalité du concept, utilisabilité du marquer , etc..
Our experiment has two goals:
\begin{enumerate}
  \item To test the operability of a system that adapts force-feedback based on
	mental workload measurement;
  \item To evaluate the influence of such a system on task performance.
\end{enumerate}
% TODO : dire le but c'est de tester notre concept et de comparer les resultats % aux pires cas (pas d'aide/des aides en permanence).
% TODO : bien dire a quoi on s'attend.
% TODO nous on a choisit d'evaluer le cas d'haptic guides.

We could make the hypothesis that the adaptation of haptic assistance based on mental workload index would help the user and would result in an increase in task performance.

\subsection{Population}
Eight participants (7 men, 1 woman) aged between 25-30 \emph{($M=26.4$, $SD=1.9$)} took part in the study. None of them had any known perception disorder. All participants were na\"ive with respect to the proposed techniques, as well as to the experimental setup and purpose.   
%il y en a qui avait deja utiliséBCI + haptique (lbonnet par exemple)

\subsection{Apparatus}
\newsavebox{\tempx}
\newsavebox{\tempy}
\begin{figure}
  \centering
  \sbox{\tempx}{\includegraphics[height=0.45\linewidth]{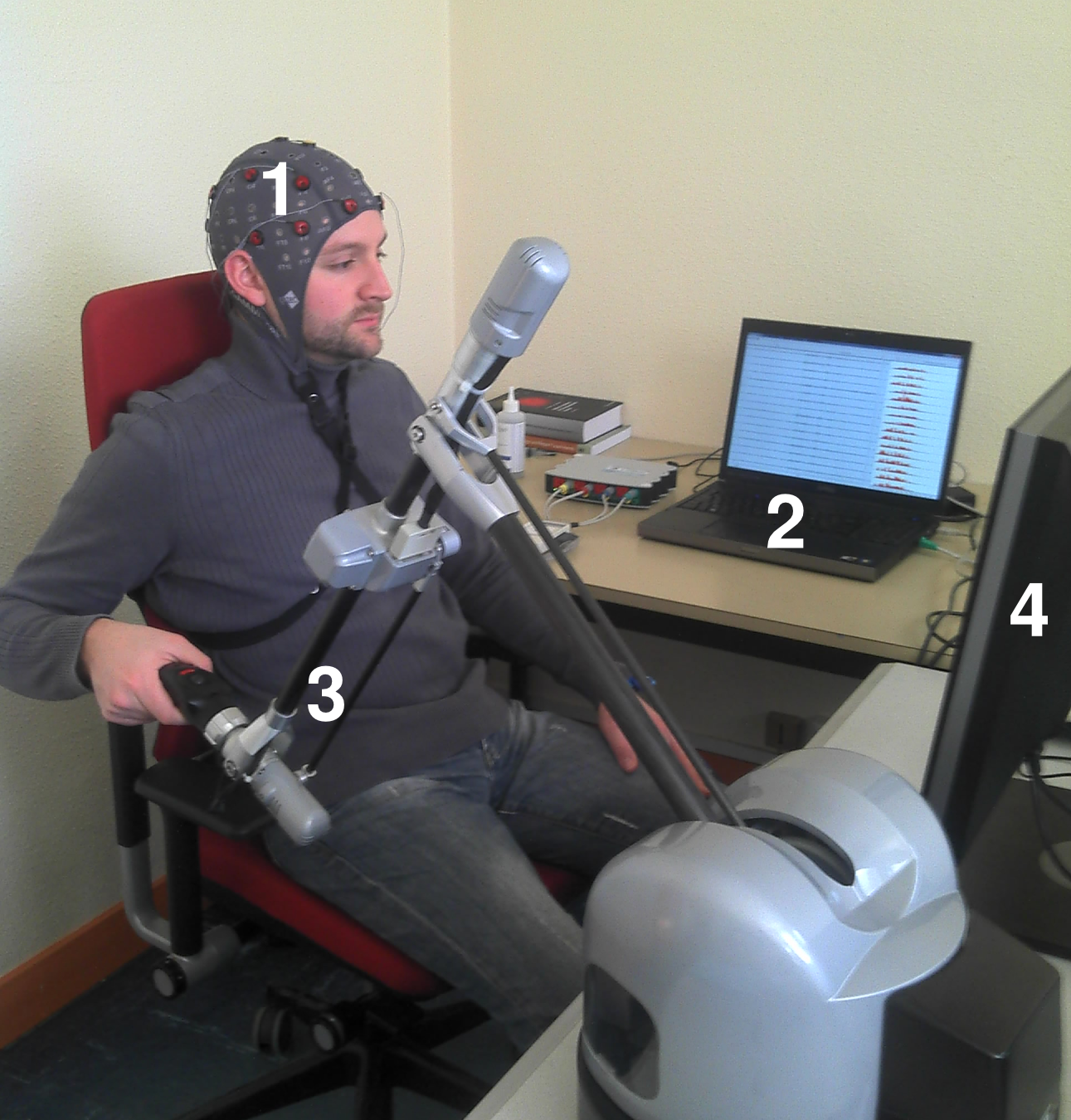}}
  \sbox{\tempy}{\includegraphics[width=0.49\linewidth]{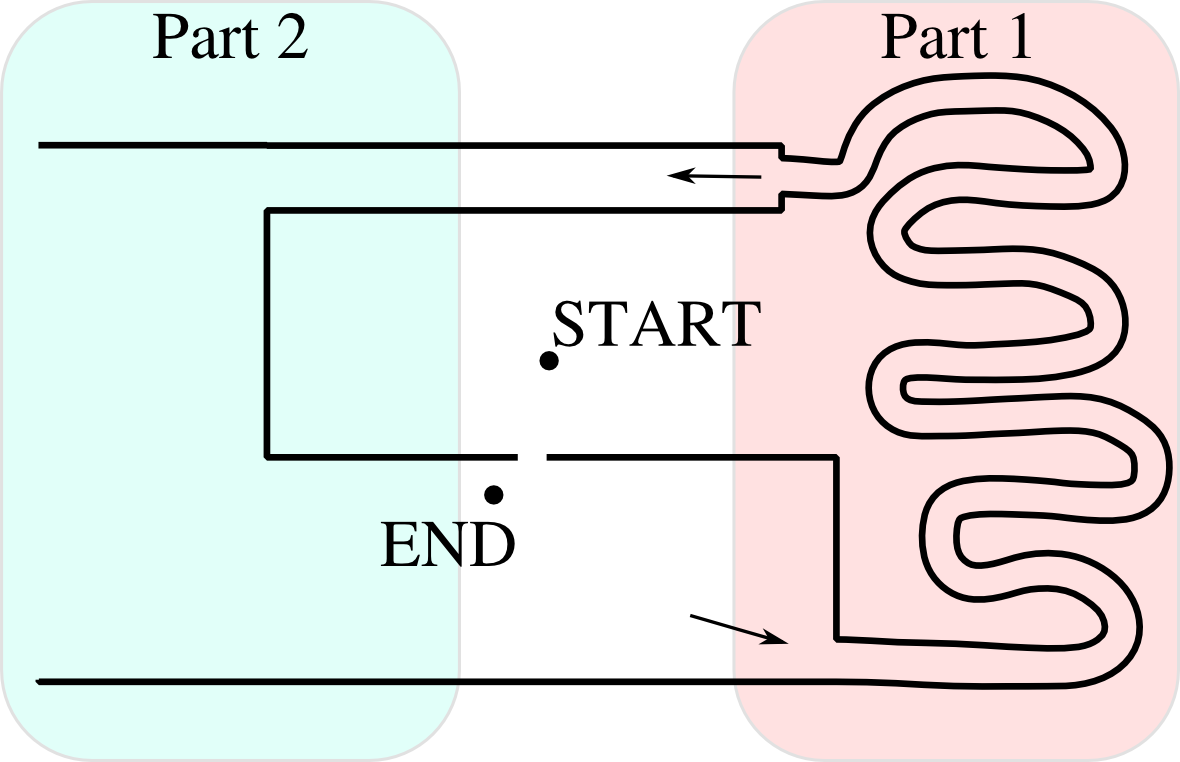}}
  \subfloat[\textbf{Apparatus} \emph{1:~EEG headset, 2:~EEG acquisition, 3:~haptic device, 4:~LCD screen where the virtual scene is displayed}.\label{subfig:apparatus}]{\usebox{\tempx}}
  \hfill
  \subfloat[\textbf{Virtual scene.\label{subfig:scene}} The Path to follow is divided into 2 parts. Part~1 (in red) is more difficult with 
  numerous turns. Part 2 (in blue) is less difficult with less turns and less borders (i.e.\ less collisions are possible).]{%
  \begin{minipage}[b]{0.49\linewidth}\centering%
  \raisebox{(\ht\tempx-\ht\tempy)/2}{\usebox{\tempy}}
  \end{minipage}
  }
  \caption{\textbf{Experimental Apparatus}}
\end{figure}
The experimental apparatus is shown in Figure~\ref{subfig:apparatus}. Participants manipulated a 2D cursor through a Virtuose 6D haptic device (Haption, Soulg\'e sur Ouette, France).
A Gtec UsbAmp was used to acquire EEG signals at a sampling rate of \SI{512}{\hertz}. EEG data were measured at positions: Fp1, Fp2, F7, F8, T7, T8, F3, F4, C3, C4, P3, P4, O1, O2, Pz and Cz according to the 10-20 international system. A reference electrode (located at FCz) and a ground electrode (located on the left ear) were also used. This electrode montage allows to cover a large surface of the scalp. Similar electrodes choices were previously used with success for recording mental workload~\cite{heger2010}.

\subsubsection{The Experimental task} consists in following a path by moving a sphere-cursor in a maze avoiding collisions (Figure~\ref{subfig:scene}). The scene is divided in two parts. These two parts aim at exhibiting two different levels of difficulty. The first part is composed of numerous turns and should lead to high difficulty, whereas the second part presents less collision possibilities to show less difficulty.

\subsubsection{The virtual environment, haptic force and collision detection} are computed and simulated with the open-source physical engine Bullet. The maze walls are composed of spherical not movable objects. % TODO: ajouter ref bullet s'il reste de la place

\subsubsection{The haptic guide} 
\begin{figure}
	\centering
	\includegraphics[width=\linewidth]{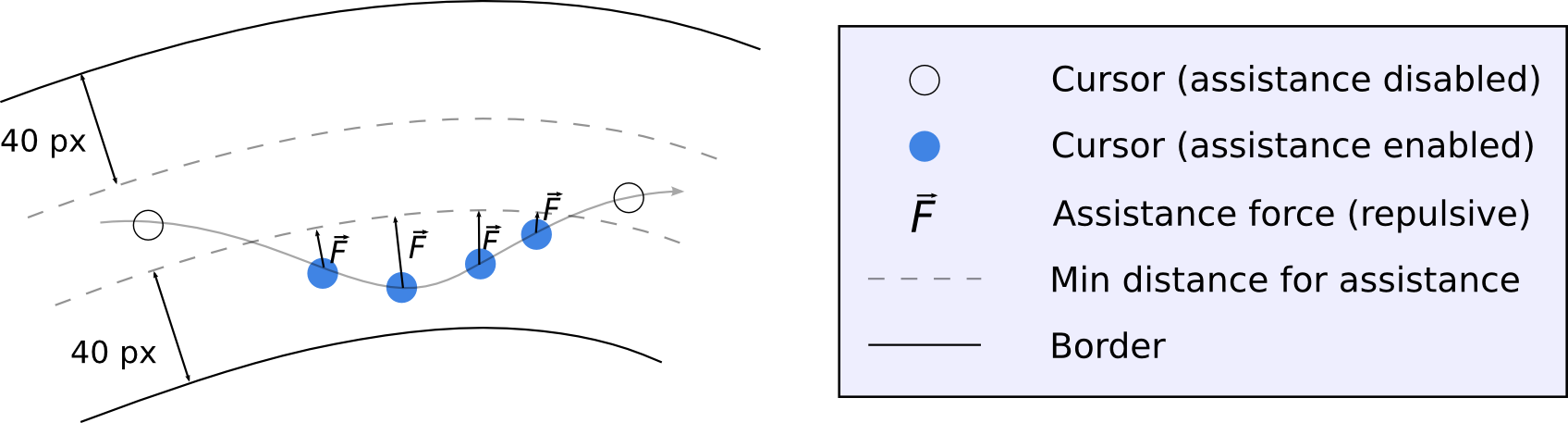}
	\caption{\textbf{Haptic assistance}. Repulsive force inversely proportional to the distance to the nearest wall: force is null if the distance to the wall is superior to 40 pixels.\label{fig:repulsion}}
\end{figure}

consists in a repulsive force inversely proportional to the distance of the 2D cursor to the nearest wall~(see Figure~\ref{fig:repulsion}). This haptic guide aims at helping the user to slide between walls avoiding collision. The cursor was colored in blue when the haptic guide was active.
%%% TODO : faire modification du dessin pour avoir le bleue seulement quand activer %% TODO
\subsubsection{A Mental workload index} is computed using OpenViBE software~\cite{openvibe}. A technique based on~\cite{hamadicharef2009} is used. EEG signals are passed through a bank of \SI{4}{\hertz} bandwidth filters centered on all the frequencies between \SI{5}{\hertz} and \SI{12}{\hertz}. A Common Spatial Pattern (CSP) method~\cite{blankertzcsp} is then used to compute spatial filters for each of them. Minimum redundancy maximum relevance feature selection is used to select the six most relevant couples of frequency band and spatial filter~\cite{mutualinforef}. A Linear Discriminant Analysis (LDA) classifier is trained on the learning data-set using a moving window of \SI{1}{\second} (overlap=\SI{0.9}{\second}).

The learning data set consists in 2 minutes of EEG activity. The first minute is recorded when the user is performing a simple control task which should lead to a low mental workload: the user has to move the cursor around a rectangle without trying to avoid collisions. The second minute corresponds to a more difficult task: the user has to move the cursor inside a spiral pattern while avoiding collision.

Online values provided by the classifier (1 for high mental workload, -1 for low mental workload) are smoothed on a \SI{1}{\second} window. The median of the last 10 values is provided to the application at a \SI{1}{\hertz} rate. This index is used to activate or inhibit the haptic assistance. 

\subsection{Procedure}
Three conditions were evaluated: \textbf{No Haptic Assistance (NO\_A)}, \textbf{Haptic Assistance based on BCI (BCI\_A)} (i.e.\ assistance activated if the mental workload is above 0), and \textbf{Haptic Assistance activated all the time (ALL\_A\@)}. Each participant performed the task in the three conditions and repeated it 3 times. To reduce order effect, the order of presentation was permuted across subjects.
%%% TODO : revoir cette phrase ..

\subsection{Collected Data}
For each trial and each participant, the mental workload index, the cursor positions and the number of collisions were recorded. The mental workload index and the cursor position were recorded at \SI{10}{\hertz}. 

At the end of the experiment, participants were asked to fill a subjective questionnaire in which they had to grade the 3 conditions according to different criteria. We used a Likert-scale where participants could rate the criteria from 1 (very bad) to 7 (very good). Criteria were: efficiency, easiness of execution and physical tiredness.
Participants were also asked to grade the correlation between their perceived mental workload and the computed mental workload index.

\subsection{Results}
%% TODO : revoir phrase pour le questionnaire !!!
\label{sec:results}
%------------------------------------------------------------
\begin{figure}
  \centering
  \subfloat[Average number of collisions.\label{subfig:collisions}]{\includegraphics[width=0.45\linewidth]{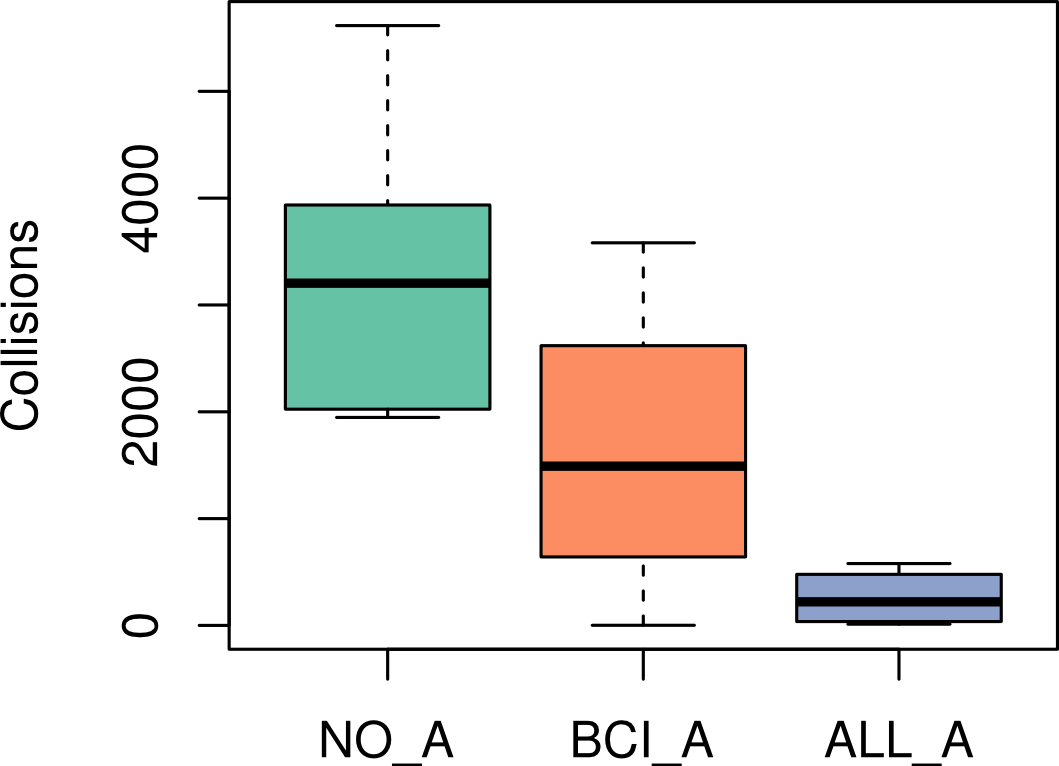}}
  \hfill
  \subfloat[Average mental workload index.%
  \label{subfig:mentalworkload}]{\includegraphics[width=0.45\linewidth]{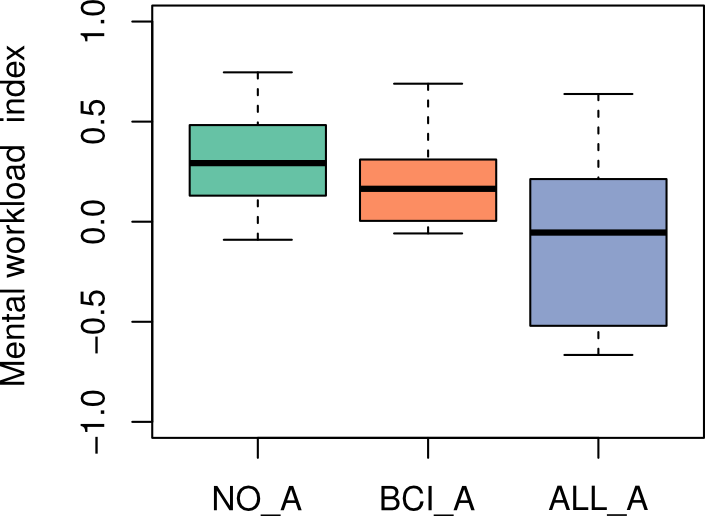}}
  \caption{\textbf{Performance and mental workload index in each condition} 
  (\emph{NO\_A:~no haptic assistance, BCI\_A:~haptic assistance activated based on BCI, ALL\_A:~haptic assistance activated all the time}). \subref{subfig:collisions} Boxplots of collisions. \subref{subfig:mentalworkload} Boxplots of mental workload index.
They are delimited by the quartile ($25\%$ quantile and $75\%$ quantile) of the distribution
of the condition over all the individuals. For each trial the median is shown.
  \label{fig:barplot}}
\end{figure}

%------------------------------------------------------------
Performances (i.e.\ number of collisions per trial) for each condition are presented in Figure~\ref{subfig:collisions}.  Friedman test shows a significant effect on assistance condition ($\chi^2=38.7, p<0.001$). Post-hoc comparisons were performed using Wilcoxon signed-rank test with a threshold of 0.05 for significance. The post-hoc analysis shows a significant difference between condition NO\_A\@ and ALL\_A\@ ($p<0.001$), and between condition NO\_A and BCI\_A ($p<0.001$). BCI\_A and ALL\_A did not differ significantly from each other ($p=0.08$).
   Activation of assistance enables to reduce the number of collisions. The average decrease over trials is $53\%$ for condition BCI\_A and $88\%$ for condition ALL\_A\@.

% maud 
%A post-hoc Wilcoxon test shows a significant difference between conditions NO\_A\@ and ALL\_A\@ ($p<.003$). 

Percentage of activation of haptic guides during trials for condition BCI\_A and each subject is shown in Table~\ref{fig:percent}. The assistance was in average activated $59\%$ of the time for condition BCI\_A ($100 \%$ for condition ALL\_A). During condition BCI\_A, assistance was activated more frequently during part~1 (64\% of the time) compared to part~2 (46\% of the time). Figure~\ref{fig:activ_point} displays the positions where the assistance was activated along the path.

%------------------------------------------------------------
\newsavebox{\tempa}
\newsavebox{\tempb}
\begin{figure}
  \centering
  \sbox{\tempa}{\includegraphics[width=0.45\linewidth]{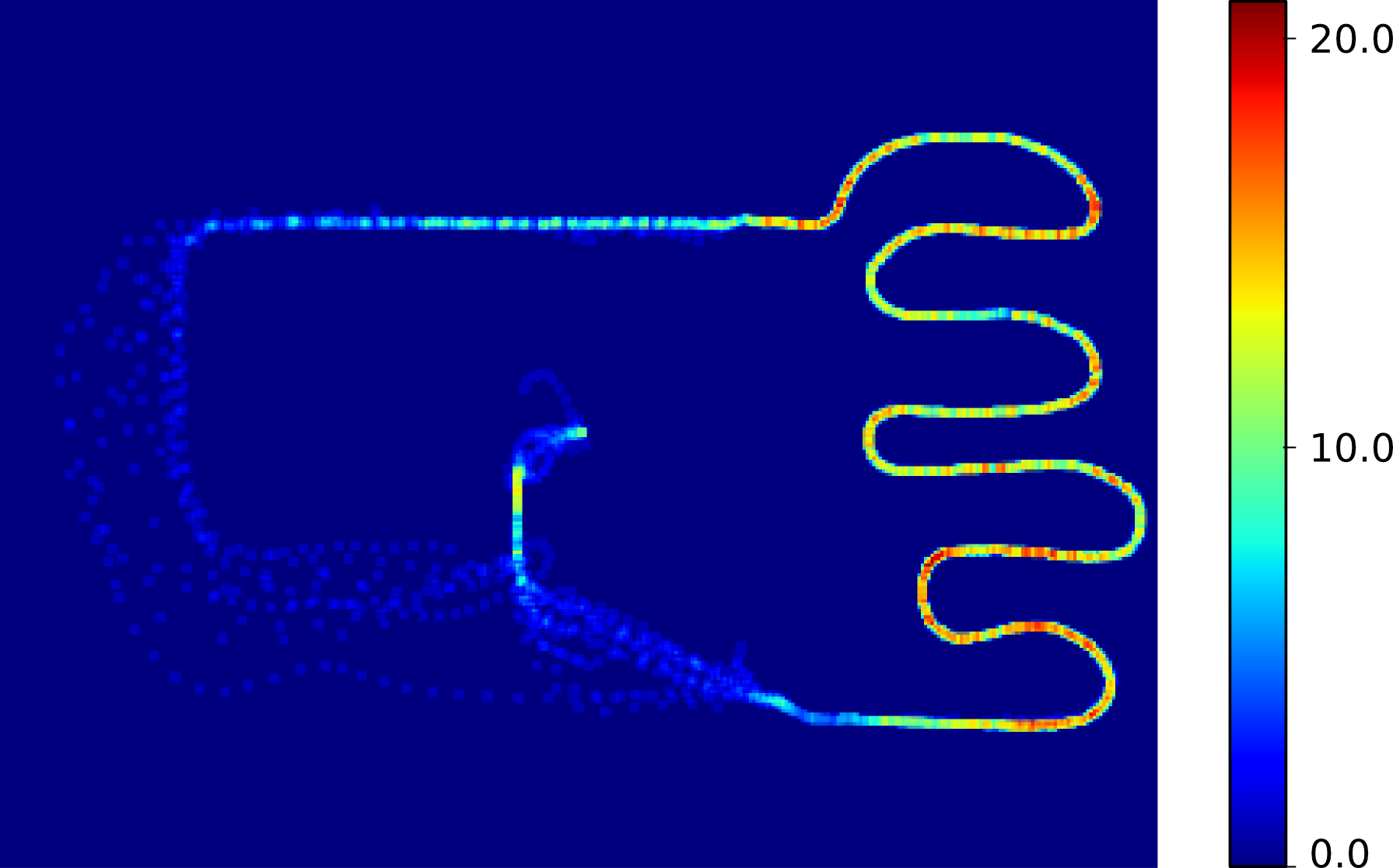}}
  \sbox{\tempb}{%
  \begin{tabular}{|l|cccccccc|cc|}
	\hline
	Subjects  & 1  & 2  & 3  & 4  & 5  & 6  & 7   & 8  & \textbf{$\mu$} & SD \\
	\hline
	\% whole  & 51 & 60 & 55 & 63 & 24 & 71 & 86  & 62 & \textbf{59} & 16.6 \\
	\hline
	\% part 1 & 54 & 62 & 62 & 57 & 26 & 90 & 98  & 65 & \textbf{64} & 20.7 \\
	\hline
	\% part 2 & 38 & 23 & 41 & 63 & 00  & 34 & 99 & 73 & 46 & 28.9 \\
	\hline
  \end{tabular}
  }
  \subfloat[][Number of activations per position.\label{fig:activ_point}]{
  \usebox{\tempa}
  }
  \hfill
  \subfloat[][Percentage of activation per subject.\label{fig:percent}]{
  \begin{minipage}[b]{0.48\linewidth}\centering%
  \centering
  \raisebox{(\ht\tempa-\ht\tempb)/2}{\usebox{\tempb}}
  \end{minipage}
  }
  \caption{\textbf{Activation of guides based on BCI (condition BCI\_A).} \subref{fig:activ_point}~Number of activation for each position for all the 8 subjects and 3 trials. \subref{fig:percent}~Percentage of activation of haptic guides during trials. 
  }
  \end{figure}
%------------------------------------------------------------

%------------------------------------------------------------
\begin{figure}
  \centering
  \includegraphics[width=\linewidth]{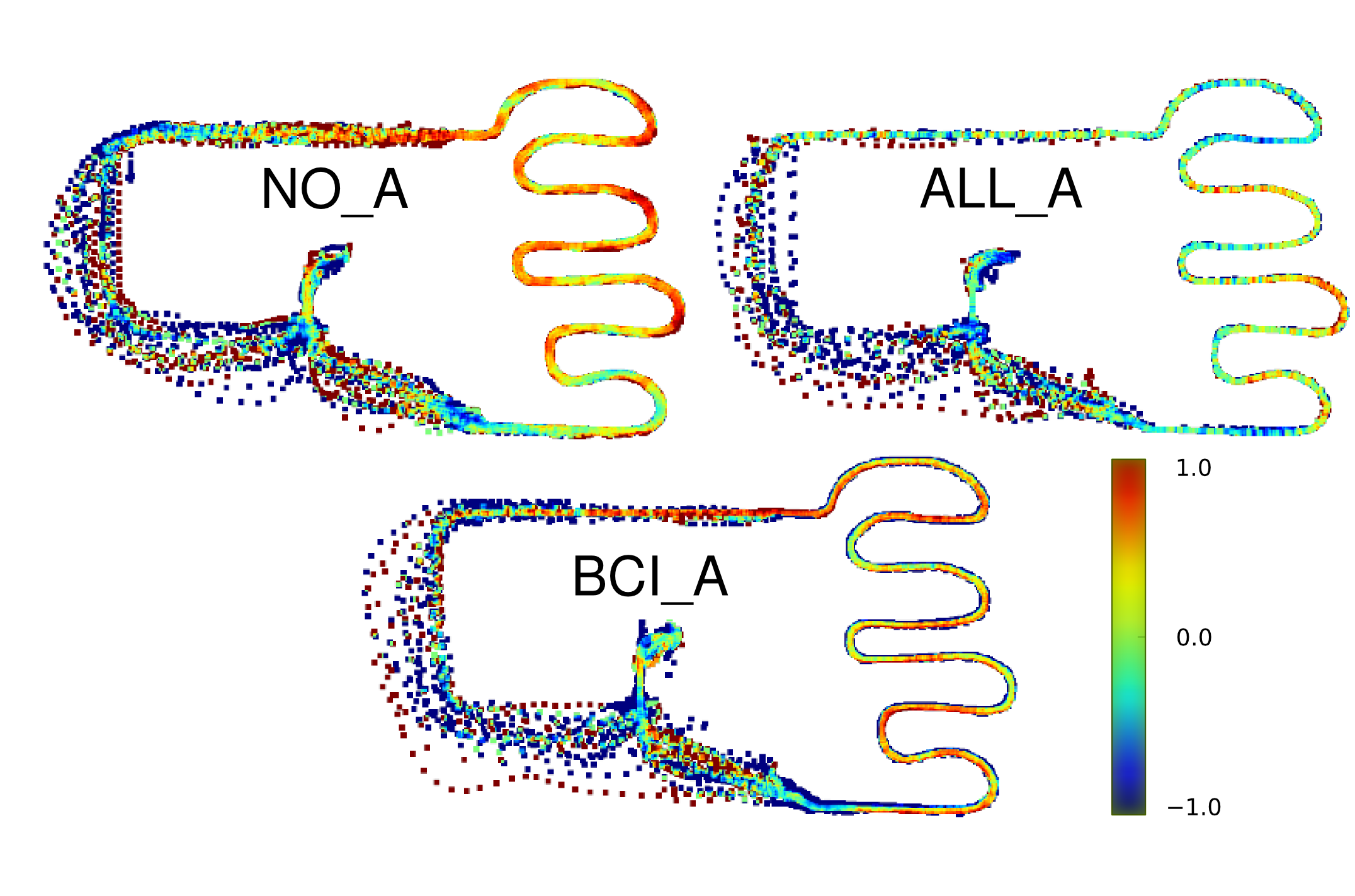}
  \caption{\textbf{Mental workload averaged over trials and subjects for each
  condition} (\emph{NO\_A:~no assistance, BCI\_A:~assistance activated based on BCI, ALL\_A:~assistance activated all the time}).
  Colored squares (9x9 pixels) are used to display mental workload index at each position
  on a 1024x768 image which represents the virtual scene. For each
  pixel, mental workload indexes were averaged over all the subjects and trials. A red color reflects a high mental
  workload index whereas a blue color corresponds to a low workload index.}
  \label{fig:path}
\end{figure}
%------------------------------------------------------------

%------------------------------------------------------------
\begin{figure}
  \centering
  \includegraphics[width=\linewidth]{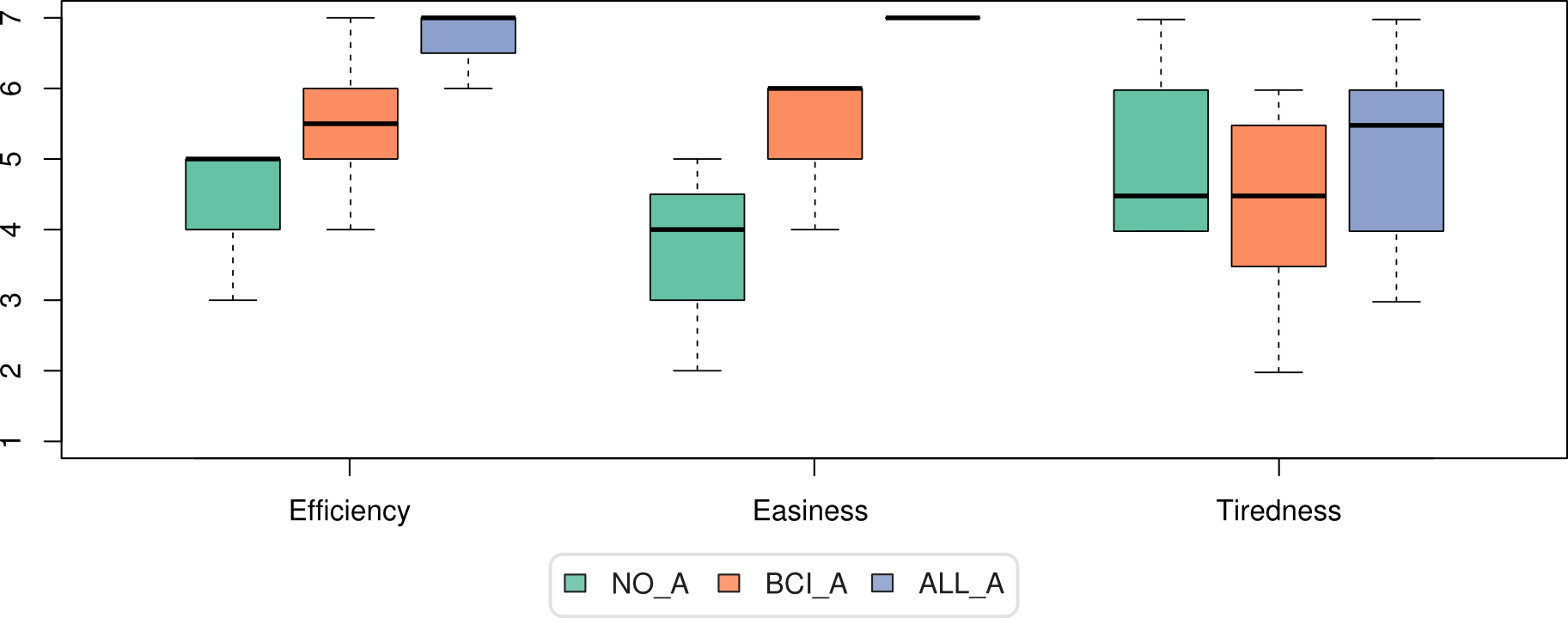}
  \caption{\textbf{Questionnaire results.} Boxplots of questionnaire reported marks.
They are delimited by the quartile ($25\%$ quantile and $75\%$ quantile) of the distribution
of the condition over all the individuals. The median is shown for each condition.
A Likert-scale (1: very bad, 7:very good) was used for criteria efficiency, easiness and physical tiredness in each condition (\emph{NO\_A:~no assistance, BCI\_A:~assistance activated based on BCI, ALL\_A:~assistance activated all the time}).\label{fig:questionnaire}}
\end{figure}
%------------------------------------------------------------

%%% anatole voudrait les valeurs moyennes pour chaque parties
%%% meme chose il voudrait qu'on reffasse toutes les analyse en regardant juste une partie (part2) puis juste part1
The computed mental workload index in each condition is presented in Figure~\ref{subfig:mentalworkload}. A Friedman test revealed a significant effect of assistance mode on the index value ($\chi^2=16.0, p<0.001$). A post-hoc analysis revealed a significant difference between condition ALL\_A and NO\_A ($p<0.001$), no significant difference between condition NO\_A and BCI\_A~($p=0.053$) and no significant difference between condition BCI\_A and ALL\_A ($p=0.21$). Mental workload index was lower when the assistance was activated.
%% TODO : thus it seems that the..

Figure~\ref{fig:path} shows the evolution of the mental workload through the path
followed by participants. During turns (part 1) we can observe a higher mental
workload index than during part 2 (Mean over subjects: $M=0.27$ for part 1, $M=-0.14$ for part 2). It is particularly clear on Figure~\ref{fig:path} for condition NO\_A.

%A Wilcoxon paired rank sum test on each trial confirmed this observation and shows a significant
%difference between part 1 and part 2 for condition NO\_A ($p <.003$) and for condition BCI\_A ($p<.006$). No significant difference was observable for condition ALL\_A ($p>.49$). 
%%% TODO : quelle est la conclusion
%

% subjective : 
Average marks of questionnaire answers are provided in Figure~\ref{fig:questionnaire}. Friedman test shows a significant effect of conditions on easiness ($\chi^2=16, p<0.001$) and efficiency ($\chi^2=14.5, p<0.001$). A higher efficiency and a higher easiness was reported when the haptic assistance was activated all the time (ALL\_A) compared to no haptic assistance condition ($p<0.001$). No significant difference was observed between BCI\_A and other conditions. This suggests that this condition is located between the two others in terms of subjective easiness and efficiency. Questionnaire results about tiredness seem to indicate that participants felt the experiment was rather tiring.

 Participants reported a high correlation between computed and felt mental workload  ( $M= 71\%$, $sd=4.7$). This suggests that the BCI system is able to provide a convincing measurement of the mental workload.

\section{Discussion}
\label{sec:discussion}
We could test the operability of our proof-of-concept system in a path-following task.  Results indicate that the proposed system works and helps the users to achieve the task. Activation of guides based on measured mental workload index allows to increase performance by significantly reducing the number of collisions. No significant difference was observed between condition ALL\_A and BCI\_A in terms of performance ($p=0.08$). This suggests that assistance activated based on BCI is almost as helpful as permanent assistance.

Results also suggest that this proof-of-concept system was able to measure a mental workload index that seems well correlated with the difficulty of the task. Indeed the measured mental workload was higher when the task was more difficult (near walls) as shown in Figure~\ref{fig:path}. Moreover the users reported a high correlation between the computed index and their perceived mental workload (above 70\%). 

In this study we used a binary adaptation (i.e.\ activating or deactivating assistance). The system could benefit from a progressive adaptation, e.g.\ more assistance if the workload is higher. We should note that 25\% of participants asked for this feature. 

A future system could also use other kinds of haptic assistance. Indeed, adapting the damping level proportionally to the user mental workload or toggling inverted damping~\cite{Williams2003} only if the user presents a high mental workload are options that should be studied. 
Concerning the measurements of the mental workload, a combination of EEG with other modalities such as Galvanic Skin Response in a multi-modal measurement system could increase the reliability of the mental workload index.
Finally, it would also be interesting to evaluate the role of BCI-based adaptation in real applications such as medical training systems notably in terms of learning performance.
\section{Conclusion}
\label{sec:conclusion}
In this paper we studied the combination of haptic interfaces and Brain-Computer Interfaces (BCI). We proposed to use BCI technology to adapt a haptic guidance system according to a mental workload index. We designed a proof-of-concept system and conducted an evaluation that revealed the feasibility and operability of such a system. High levels of mental workload could be well identified by the BCI system in the most difficult parts of a path-following scenario. Toggling haptic guides only when the user presented a high mental workload could improve task performance indicators. Taken together our results pave the way to novel combinations of BCI and haptics. Haptic feedback could be fine-tuned according to various mental states of the user, and for various purposes. Future work will focus on the extraction of other mental states, and on real applications such as haptic-based medical simulators.

\section{Acknowledgements}
This work was supported by the French National Research Agency within the OpenViBE2 project and grant ANR-09-CORD-017. The authors would also like to thank all the participants who took part in our study for their patience and kindness.

\bibliographystyle{splncs}
\bibliography{mybib}

\end{document}